\documentstyle[12pt,epsfig]{article}

\newcommand{\be}{\begin{equation}}
\newcommand{\ee}{\end{equation}}
\newcommand{\ba}{\begin{eqnarray}}
\newcommand{\ea}{\end{eqnarray}}
\newcommand{\bb}{\begin{thebibliography}}
\newcommand{\eb}{\end{thebibliography}}
\newcommand{\ci}[1]{\cite{#1}}
\newcommand{\bi}[1]{\bibitem{#1}}
\newcommand{\lab}[1]{\label{#1}}

\begin{document}
\sloppy
\thispagestyle{empty}

\mbox{}
\vspace*{\fill}
\begin{center}
{\LARGE\bf Effects of Pomeron Coupling in  } \\
\vspace{2mm}
{\LARGE\bf  Diffractive Reactions}\\

\vspace{2em}
\large
S.V.Goloskokov
\\
\vspace{2em}

{
\it
Bogoliubov Laboratory of Theoretical  Physics,
  Joint Institute for Nuclear Research.}
 \\

{
\it
Dubna 141980, Moscow region, Russia.}\\
\end{center}
\vspace*{\fill}
\begin{abstract}
\noindent
The diffractive 2-jet production reactions in the proton-proton and
lepton-proton processes are discussed.  It is shown that they may be very
suitable for studying the properties of the pomeron couplings with quarks
and hadrons in future polarized experiments at HERA, especially for
HERA-$\vec N$ project.
 \end{abstract}
 \vspace*{\fill}
 \newpage %

\section{Introduction}
\label{sect1}

Recently, the interest has increased in investigating polarized
processes. To study the spin properties of the
hadron interaction at high energies, the precise information about
different spin asymmetries is important. For such experiments  it is
necessary to have high energy and high intensity polarized beams. The only
project which is now being realized is the RHIC spin program \ci{bunce}.
Physics with a polarized beams is discussed now for the HERA
accelerator (see \ci{schaefer}).  An important part of
this program is the HERA-$\vec N$ project \cite{now} with the fixed
polarized target on the proton beam at HERA.
At the 'phase I' (with unpolarized proton beam)  the HERA-$\vec N$
experiment will allow one to study different single-spin asymmetries in the
nucleon-nucleon reaction.
 When the polarized protons at HERA become available,  the
HERA-$\vec N$ experiment at the 'phase II' will permit one to study
double-spin asymmetries. As a result, a great deal of information on the
spin structure of QCD can be obtained.

The diffractive events with a large rapidity gap in deep inelastic
lepton--proton scattering
have recently been studied  in the H1 and ZEUS experiments at HERA
\ci{gap1,gap2}.  The natural explanation of these events is based on
the hard photon--pomeron interaction. This permits one to obtain new
information about the structure of the pomeron and its couplings.
So, the HERA accelerator is the best places to study the pomeron
properties.

If the polarized program at HERA is realized,
 the diffractive effects with polarized particles can be
investigated. Then, the question about the spin structure of the
pomeron arises.  This problem is very important for the following
reasons:
\begin{itemize}
\item{There are many observations of not small spin effects at high
energies and fixed momentum transfer \ci{nur}.
For example, the single-spin transverse asymmetry for $|t| \ge 1.5(GeV)^2$
and $\sqrt{s}=(19-24)GeV$ is about 10-20\% \ci{pol} and may be independent
of energy.}
\item{Attempts to extract the spin-flip amplitude from  experimental
data \ci{akch} show that the ratio of spin-flip and spin-non-flip
amplitudes can be about 0.1-0.3 at high energies.}
\item{Some model approaches predict the same  ratio of spin-flip and
spin-non-flip amplitudes in the $s \to \infty, |t|/s
\to 0 $ limit (see \ci{gol,soff} e.g.).}
 \end{itemize}
Just in all these cases the pomeron exchange should give a contribution.
Thus, the pomeron might have a complicated spin structure.

The pomeron is a colour--singlet vacuum $t$-channel exchange that
can be regarded
as a two-gluon state \ci{low}.
The pomeron contribution to the hadron high energy amplitude can be
written
as a product of two pomeron vertices $V_{\mu}^{hhI\hspace{-1.1mm}P}$
multiplied by some function $I\hspace{-1.6mm}P$ of the pomeron.
As a result,  the quark-proton high-energy amplitude
looks like
\be
T(s,t)=i I\hspace{-1.6mm}P(s,t) V_{hhI\hspace{-1.1mm}P}^{\mu}(t) \otimes
V^{hhI\hspace{-1.1mm}P}_{\mu}(t).    \label{tpom}
\ee

In the nonperturbative two-gluon exchange model  \cite{la-na} and the
BFKL model \cite{bfkl} the pomeron couplings have a simple matrix
structure:
\be
V^{\mu}_{hh I\hspace{-1.1mm}P} =\beta_{hh I\hspace{-1.1mm}P}
\gamma^{\mu},
\label{pmu}
\ee
 which leads to spin-flip effects decreasing with energy as a power of
$s$.
We call this form the standard coupling.

The situation does change drastically when large-distance loop
contributions are considered. As a result, the spin structure
of the pomeron coupling becomes more complicated. These effects can be
determined by the hadron wave function for the pomeron-hadron couplings or
by the gluon-loop $\alpha_s$ corrections for the quark-pomeron coupling.

Pomeron-proton coupling is mainly connected with the proton structure at
large distances.  The perturbative calculation of this coupling is difficult.
Moreover, for a momentum transfer of about few $GeV^2$ the
nonperturbative contributions are important. Some
models can be used to study the spin structure of the
pomeron-proton coupling. For example, the diquark model \ci{kroll} takes
effectively into account the nonperturbative contributions. It can leads
to the spin--flip in the pomeron--proton vertex \ci{kopel}.

In models  \cite{gol,soff} the spin-flip effects do not vanish
as $s \to \infty$.  It has been shown \cite{gol} that the pomeron--proton
vertex is of the form
\be
V_{ppI\hspace{-1.1mm}P}^{\mu}(p,r)=m p^{\mu} A(r)+ \gamma^{\mu} B(r),
\label{prver}
\ee
where $m$ is the proton mass, the amplitudes $A$ and $B$ are connected
with the proton wave function.
The model \cite{gol} predicts that the ratio
\be
m^2 |A|/|B|\sim 0.2
\ee
at $|t| \ge 1GeV^2$, which leads to the weak energy dependence of single
and double spin transverse asymmetries which are about $10 - 15\%$. The
discussion of this question can be found in \ci{sel}.

The form of the quark-pomeron coupling
 $V_{qqI\hspace{-1.1mm}P}^{\mu}$
has been studied in \cite{gol-pl}.
It was shown that besides the standard pomeron vertex
(\ref{pmu}) determined by the diagrams, where gluons interact
with one quark in the hadron  \cite{la-na}, the large-distance
gluon-loop effects should be important.
 The perturbative calculations \ci{gol-pl} give the following
form of this vertex:
\be
V_{qqI\hspace{-1.1mm}P}^{\mu}(k,r)=\gamma^{\mu} u_0+2 M_Q k^{\mu} u_1+
2 k^{\mu}
/ \hspace{-2.3mm} k u_2 + i u_3 \epsilon^{\mu\alpha\beta\rho}
k_\alpha r_\beta \gamma_\rho \gamma_5+i M_Q u_4
\sigma^{\mu\alpha} r_\alpha,    \label{ver}
\ee
where $k$ is the quark momentum, $r$ is the momentum transfer and  $M_Q$
is the quark mass.  So, in addition to the $\gamma_\mu$ term, the new
structures immediately appear from the loop diagrams.  The functions
$u_1(r) \div u_4(r)$ are proportional to $\alpha_s$. These
functions can reach $30 \div 40 \%$ of the standard pomeron term
$u_0(r)$ for $|r^2| \simeq {\rm Few}~ GeV^2$ \cite{golsel}.

The new form of the pomeron--quark coupling (\ref{ver}) should modify
various  spin
asymmetries in high--energy diffractive reactions \cite{klen,golpr,golall}.
 In this report we shall discuss the single and double spin asymmetries
in  polarized diffractive reactions, which may be studied
 in the future HERA-$\vec N$ project to test the spin structure of the
 pomeron couplings.
\section{Spin Asymmetries}
\label{sect2}
The single-spin asymmetry which can be studied at the 'phase I' of
HERA-$\vec N$ experiment is determined by the relation
\be
A_{\perp} =\frac{\Delta \sigma}{\sigma}=
\frac{\sigma(^{\uparrow})-\sigma(^{\downarrow})}
{\sigma(^{\uparrow})+\sigma(^{\downarrow})}.
\ee

We shall discuss here the single transverse spin asymmetry in the
$p\uparrow p \to p+Q \bar Q+X$ process.  This reaction is determined by
the pomeron exchange between the proton and the produced $Q \bar Q$ pare at
high energies and small $x_p$ ($x_p$ is a fraction of the initial proton
momentum carried off by the pomeron).
The distributions of the
cross sections $\sigma$ and $\Delta \sigma$  over $p_{\perp}^2$ of jets
can be written in the form
\be
\frac{d \sigma(\Delta \sigma)}{dx_p dt dp_{\perp}^2}=\{1,A^h_{\perp}\}
\frac{\beta^4 |F_p(t)|^2 \alpha_s}{128 \pi s x_p^2}
\int_{4p_{\perp}^2/sx_p}^{1} \frac{dy g(y)}{\sqrt{1-4p_{\perp}^2/syx_p}}
\frac{ N^{\sigma(\Delta \sigma)}
(x_p,p_{\perp}^2,u_i,|t|)}{(p_{\perp}^2+M_Q^2)^2}. \lab{si}
\ee
Here $g$ is the gluon structure function
of the proton, $p_{\perp}$ is a transverse momentum of jets, $M_Q$
is a quark mass, $N^{\sigma(\Delta \sigma)}$ is a trace over the quark loop,
$\beta$ is a pomeron coupling constant,
and $F_p$ is a pomeron-proton form factor.
In (\ref{si}) the coefficient equal to unity appears in $\sigma$ and
the transverse hadron asymmetry $A^h_{\perp}$ at the pomeron-proton vertex
appears in $\Delta \sigma$.
This asymmetry $A^h_{\perp}$ is determined by the interference between the
amplitudes $A$ and $B$ in the pomeron--proton coupling (\ref{prver}).
\be
A^h_{\perp} \simeq \frac{2m \sqrt{|t|} \Im (AB^{*})}{|B|^2}.
\label{epol}
\ee

We use a simple form of the gluon structure function
\be
 g(y)=R (1-y)^5/y,\;\;\;\;R=3.   \lab{gy}
 \ee

In the diffractive--jet production the main contribution
is determined by the region where the quarks
in the loop are not far from the mass shell.
So, we can assume that the asymmetry $A^h_{\perp}$ (\ref{si}) can be
determined by the soft pomeron, and it coincides with the elastic
transverse hadron asymmetry. In our further estimations we use the
magnitude $A^h_{\perp}=0.1$. Some details of calculations can be found in
\ci{golpr}.

 It has been found that both $\sigma$ and $\Delta \sigma$ have a similar
dependence $ \sigma(\Delta \sigma) \propto 1/x_p^2$ at small $x_p$. This
allows one to study asymmetry at small $x_p$ where the pomeron exchange
is predominated because of a high energy in the quark-pomeron system.

Our predictions for the asymmetry $A_{\perp}$
at $\sqrt{s}=40GeV$, $x_p=0.05$ and $|t|=1GeV^2$
for a standard quark-pomeron vertex
(\ref{pmu}) and a spin-dependent quark-pomeron vertex (\ref{ver}) are shown
in Fig.1 for light-quark jets.
It is easy to see that the shape of asymmetry is different for standard and
spin-dependent pomeron vertices. In the first case it is approximately
constant and in the second it depends on $p^2_{\perp}$.   The estimated
errors for the integrated luminocity 240 $pb^{-1}$ are shown in Fig.1 too.
The structure of the quark-pomeron vertex can be studied from the
$p^2_{\perp}$ distribution of single-spin asymmetry (the
region with the errors smaller than 1\% is $1Gev^2 <p^2_{\perp}< 10GeV^2$).
So, from this asymmetry the structure of the quark-pomeron vertex can be
determined.

 The cross sections $\sigma$ and $\Delta \sigma$ integrated over
 $p^2_{\perp}$ of jet have been calculated too. The asymmetry obtained
from these integrated cross sections does not practically depend on the
quark-pomeron vertex structure.  It can be written in both the cases in
the form
\be
A1=\frac{\int dp^2_{\perp} \Delta  \sigma}{\int dp^2_{ \perp}\sigma} =
0.5\;A^h_{\perp}  \lab{a1}
\ee
Thus, the information on the transverse hadron asymmetry $A^h_{\perp}$ at
the pomeron-proton vertex can be obtained from the integrated asymmetry
(\ref{a1}).

At the second phase of the HERA-$\vec N$ project, double--spin asymmetries
can be studied.
We shall discuss here the longitudinal double spin asymmetry determined by
the relation
\be
A_{ll}=
\frac{\Delta \sigma}{\sigma}=\frac{
\sigma(^{\rightarrow} _{\Leftarrow})-\sigma(^{\rightarrow} _{\Rightarrow})}
{\sigma(^{\rightarrow} _{\Rightarrow})+\sigma(^{\rightarrow} _{\Leftarrow})}.
\lab{asydef}
\ee
For the spin-average and longitudinal polarization of the proton beam
the $B$ term in (\ref{prver}) is predominant. As a result, the longitudinal
double spin asymmetry does not depend on the pomeron-proton vertex structure.

The $A_{ll}$  asymmetry of the  $Q \bar Q$ production in the $pp$
diffractive reaction is proportional to the ratio \ci{golall}
\be
 C_g= \frac{\Delta g}{R}.
\ee
Here  $ \Delta g$ is  the first moment
of $ \Delta g(y)$
\be
 \Delta g =\int_{0}^{1} dy \Delta g(y), \lab{i3}
\ee
which is unknown now, and R is determined in (\ref{gy}).
To solve the proton spin crisis \ci{efr} a large magnitude of
 $ \Delta g \sim 3$ is important.
This leads to $C_g \sim 1$ which will be used in here.
However, the magnitude  $\Delta g \sim 1$ may be more preferable now
\ci{rams}.  Then, the resulting asymmetry will
decrease by a factor of 3.

Our predictions for the $A_{ll}$ asymmetry at $\sqrt{s}=40GeV$ (HERA-N
energy) for light and heavy (C) quark production can be found in
 \ci{golall}.  The magnitude of the obtained asymmetry strongly depends on
the structure of the quark-pomeron vertex but the shapes of these curves
are similar.  For a spin-dependent quark-pomeron vertex the $A_{ll}$
asymmetry is smaller by a factor of 2 because $\sigma$ in (\ref{asydef})
is larger in this case due to the contribution of other $u_i$ structures.
Therefore, the knowledge of the proton spin-dependent structure function
$\Delta g$ is necessary to study the pomeron spin structure from the
$A_{ll}$ asymmetry in $pp$ polarized diffractive reactions.

The very important test of the
pomeron coupling can be performed in polarized lepton--proton reactions.
Really, in this case we know explicitly the lepton part of the interaction.
As a result, the spin asymmetries in diffractive $Q \bar Q$ production
will depend only on the structure of the pomeron couplings with
the quark and proton.
Here, we shall discuss the double--spin longitudinal asymmetry in the
reaction $e+p \to e'+p'+Q \bar Q$ \ci{f2}.

The asymmetry in the diffractive $Q \bar Q$ production is shown in Fig. 2.
It is found that
the asymmetry for the standard quark--pomeron vertex is very simple in
form
\be
A_{ll}=\frac{y x_p (2-y)}{2-2y+y^2}.
\ee
There is no any $k_\perp$ and $\beta$ dependence here. For the
spin--dependent pomeron coupling  the $A_{ll}$  asymmetry is smaller than
for  the standard pomeron vertex and depends on $k^2_{\perp}$. Thus, one
can use the $A_{ll}$ asymmetry to test the quark-pomeron coupling
structure.  Note that  $\Delta \sigma$  in (\ref{asydef}) can be used to
determined the diffractive contribution to the $g_1$ spin--dependent
structure function can be determined \ci{g1g}.  The obtained low -$x$
behaviour of $ g_1(x)$ has a singular form like $1/(x^{0.3} \ln^2(x))$
which is compatible with the SMC data for $g^p_1(x)$ \ci{smc}.
\section{Conclusion}
\label{sect3}
In this report we have discussed the possibility to test the pomeron
coupling structure in diffractive reactions. It is shown that the
diffractive 2-jet production may be very suitable for this purpose.
In all the cases the distribution of the asymmetries over the transverse
jet momentum is very sensitive to the quark-pomeron vertex structure. For
the standard pomeron vertex the asymmetry is practically independent of
$k^2_{\perp}$ of jet. Otherwise, it should have a very definite
$k^2_{\perp}$--dependence.

The information about the proton-pomeron coupling can be obtained from the
single-spin transverse asymmetry integrated over the transverse momentum
of jets. Moreover, one can understand if this diffractive pomeron
coincides with the pomeron in elastic processes.

To study spin asymmetries with the transversely polarized protons, it is
necessary to register the recoil hadron. Thus, it is important to have
the RECOIL hadron detector. If this problem is solved  at HERA-$\vec
N$, the single and double transverse asymmetries ($A_{\perp}$ and
$A_{l\perp}$) in the diffractive  $Q \bar Q$ production reactions may be
very useful to study the spin structure of the pomeron-proton coupling.
Relevant asymmetries in the diffractive vector meson production can be used
too. These reactions may be easy in detection of the final state but
they have a smaller cross section. This should increase errors in
asymmetries.

It is very important from the experimental point of view that the
investigated here diffractive 2-jet production reactions have a large cross
section. This permits one to study single and double spin asymmetries in
lepton-proton and proton-proton reactions with the accuracy less than 1\%
at the integrated luminosity about 200 $pb^{-1}$. Thus, these reactions
may be used to study the pomeron couplings in future polarized experiments
at HERA and HERA-$\vec N$.

\newpage
  \vspace*{-.5cm}
\epsfxsize=10cm
\centerline{\epsfbox{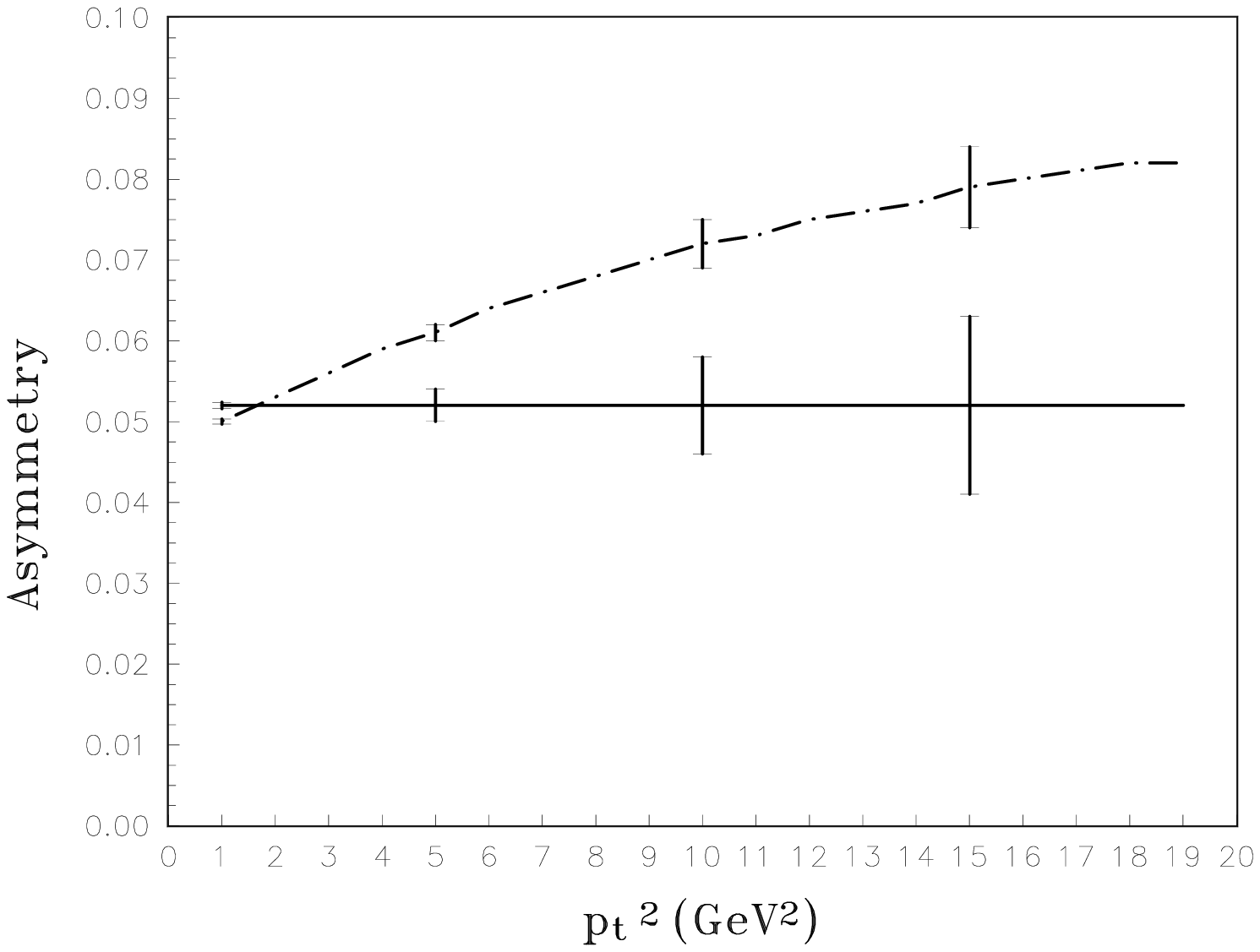}}
  \vspace*{.2cm}
\begin{center}
Fig.1 ~$p^2_{\perp}$--dependence of $A_{\perp}$ asymmetry and the estimated
   errors. Solid line -for standard;
   dot-dashed line -for spin-dependent quark-pomeron vertex.
\end{center}

  \vspace*{-.1cm}
       \hspace*{.5cm}
\epsfxsize=10cm
{\epsfbox{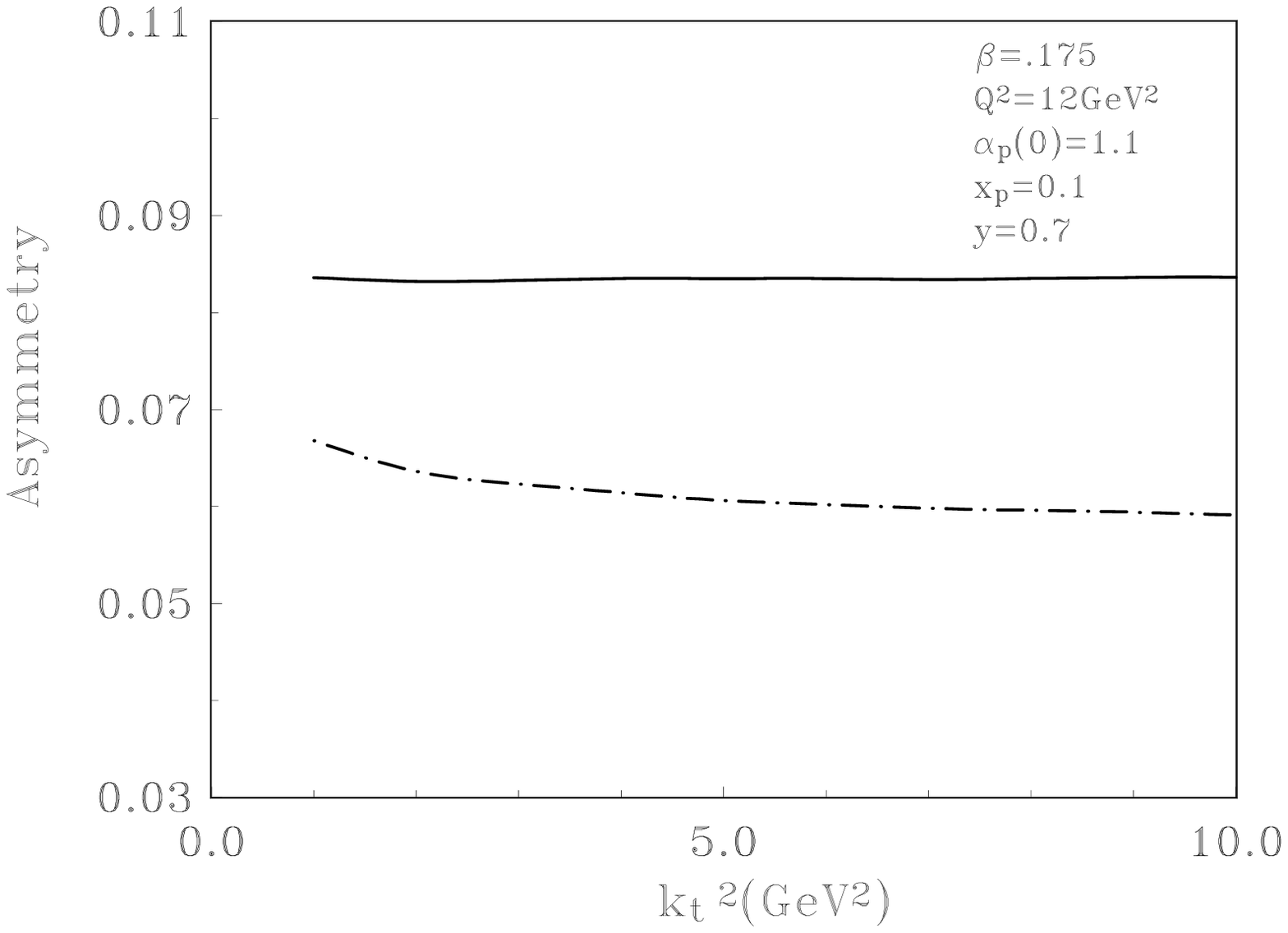}}
  \vspace*{.3cm}
\begin{center}
Fig.2 ~$k^2_{\perp}$-- dependence of $A_{ll}$ asymmetry.
Solid line -for the standard vertex;
dot-dashed line -for the spin-dependent quark-pomeron vertex.

\end{center}

\end{document}